# Second-Harmonic Magnetoacoustic Ultrasound from Magnetic Nanoparticles under Radiofrequency Electromagnetic Fields

R. Marqués-Gómez [1,2,5], J. Melchor [1,5,6,7], A.C. Moreno Maldonado [2,3], C. Marquina [2,3], G. Goya [2,3], M. R. Ibarra [2,3,8], G. Rus [4,5,6,7]

*1 Department of Statistics and Operational Research, University of Granada, Spain*

*2 Instituto de Nanociencia y Materiales de Aragón (INMA), CSIC-Universidad de Zaragoza, Zaragoza 50009, Spain*

*3 Departamento de Física de la Materia Condensada, Universidad de Zaragoza, Zaragoza 50009, Spain*

*4 Department of Structural Mechanics and Hydraulic Engineering, University of Granada*

*5 Ultrasonics Lab., University of Granada, Spain*

*6 Instituto de Investigación Biosanitaria, ibs.GRANADA, 18012 Granada, Spain*

*7 Research Unit "Modelling Nature" (MNat), Universidad de Granada, 18071 Granada, Spain*

*8 Faculty of Space Technologies, AGH University of Science and Technology, Krakow, Poland*

## Abstract

We report experimental evidence of mechanical wave generation by magnetic nanoparticles (MNPs) under radiofrequency electromagnetic fields (EMFs, 65mT, 800 kHz, at 100ms bursts) in the ultrasonic (US) frequency range. We developed an experimental setup to detect the second harmonic (SH) signal under isothermal conditions. This SH-US signal was observed only in samples containing MNPs and consequently originated from the EMF interaction on samples containing MNPs. These findings suggest that US generation due to magneto-acoustic interaction in the MNPs could be responsible for the cell damage induced by EMF in cells containing MNPs without increasing the temperature. Experiments performed on magnetically aligned MNP clusters in gelatin significantly enhance the amplitude of the SH signal compared to non-aligned samples. This phenomenon could pave the way for new theoretical and practical approaches as for "in vivo" magnetoacoustics theragnostic.

The utilization of MNPs in biomedical and clinical procedures for diagnostic and therapeutic purposes (theragnostic) has steadily increased. One of these applications is magnetic fluid hyperthermia (MFH), a cancer treatment designed to transfer heat from locally injected MNPs to cancer cells [1]. These MNPs serve as nanoheaters that absorb energy from an externally applied EMF and generate heat, to raise the local temperature and to kill malignant cancer cells (42-46 ºC). This energy absorption by MNPs under EMF is the source of the heating properties used for MFH [2, 3]. Over the last few years, it has been shown that the energy absorption by MNPs produces physical damage on biological membranes as a consequence of the energy released by the MNP, which cannot be explained by considering only the effects of temperature rise [4]. Asín et al. investigated the parameters of controlled cell death under MFH conditions and concluded that temperature increase alone could not be responsible for the cell death, indicating the involvement of an underlying no thermal process [5,6].

Carrey et al. [7] first proposed the hypothesis that mechanical oscillations of

MNPs under an EMF gradient could account for the observed results. By investigating the forces acting on MNPs in a non-homogeneous EMF, this work highlighted the crucial role played by the magnetic hysteresis cycle in driving the oscillation of the particles at a frequency twice that of the applied EMF. Kellnberger et al. [8], proposed a theoretical analysis of the pressure generation due to MNPs under an EMF. This model considers the internal energy of the magnetic system and obtains an expression for the local pressure generated by the second harmonic of the applied EMF under adiabatic conditions. They carried out an experiment using an optical interferometric sensor that confirmed the occurrence of the pressure generation effect at both the nanoscale and macroscopic levels. Guo et al. [9] extended this theoretical framework to predict the generation of acoustic waves in MNPs under the influence of a continuous EMF wave, providing additional support to the theoretical framework with experimental validation using a laser vibrometer. They already established the relevance of either the continuous or discontinuous applied EMF to raise the temperature (as in MFH) and generate SH acoustic waves. Hu et al. and Mariappan et al. [10,11] have investigated the employment of MNPs as contrast agents in magnetoacoustic tomography by magnetic induction, demonstrating the potential of MNPs in the formation of US images produced by MNPs by applying magnetic pulses. However, these studies are limited because they focus on detection of the fundamental frequency (first harmonic) of the applied EMF.

Our study confirms for the first time the theoretical predictions by Carrey, that SH-US is generated under isothermal conditions, which provides the basis for understanding cell death in response to EMF without an increase in temperature. The observation of SH is intrinsically linked to the magnetic properties of MNPs and depends on the presence of a magnetic field gradient. We have developed an experimental setup to detect second harmonic US generation by MNPs under EMF. A key advantage of our proposed method lies in the use of piezoelectric transducers, already used in clinical instruments. The experiments were performed at a frequency $f = 800\ kHz$ and magnetic flux densities ($\Phi$) up to $65\ mT$. In order to avoid the increase of temperature ($\Delta T < 1$ °C), we applied 100 ms burst of the RF EMF. The signal detection methodology was optimized to minimize the effects of electrical noise generated by the EMF excitation. This optimization included careful calibration and tuning of the equipment to ensure accurate measurement of the acoustic wave signal while reducing interference from external sources.
We observed the presence of a signal at the second harmonic frequency (1.6 MHz) in the samples containing MNPs. These findings constitute a relevant US-generation fingerprint due to magneto-acoustic interaction in the MNPs. Our methodology represents a novel advancement in MNP-based technologies, providing a reliable and precise means of measuring the acoustic response of magnetic systems under EMF. This knowledge could be instrumental in developing new biomedical applications, potentially enhancing the performance and effectiveness of MNP-based technologies. Additionally, the US generation observed in this study may help to explain some of the non-thermal effects seen in MNP-loaded cells under MFH [5,6]. Efficient targeting of MNPs also opens up the possibility of performing US theragnostics by combining acoustic imaging with MFH treatment and drug delivery [12].
Theoretical predictions [7] suggest that magnetoacoustic interactions in MNPs can enhance SH-US generation when a uniform magnetic field is applied in addition to an EMF gradient. However, experimental validation remains challenging due to the inherent difficulty in achieving precise spatial alignment among: the EMF generator, the ultrasonic (US) detector system, and any external magnetic sources. To overcome this limitation, we applied a uniform static magnetic field during the gelation process of the samples, thereby inducing a preferential alignment of the magnetic nanoparticles (MNPs) along the field direction. This strategy facilitates the fabrication of anisotropic MNP-embedded samples with defined magnetic orientation, enabling systematic measurements of the second harmonic ultrasonic (SH-US) response under conditions where the applied EMF and its spatial gradient are oriented either parallel or perpendicular to the pre-established magnetic alignment.
Magnetic nanoparticles of $MnFe_2O_4$, with an average size of (37 ± 7) nm, were synthesized using the co-precipitation method as described in [1]. A gelation process was followed to align the MNPs in a rigid medium. Commercial ROYAL® gelatin at a 10% concentration and Getzan® gelatin at an 8% concentration were utilized. In a 50 mL falcon tube, 250 µL of MNPs and 1 mL of gelatin were added, with ROYAL gelatin mixed at 60°C and Getzan® gelatin at 95°C.
The mixture was thoroughly mixed by pipetting and then sonicating in a 60°C water bath. After confirming uniform dispersion and the absence of bubbles, it was rapidly cooled at -20°C for 5 minutes to prevent MNP precipitation and agglomeration. A portion of the resulting nanoparticle-loaded hydrogel was pipetted into the bottom of Falcon tubes. To align the MNPs, the Falcon tubes were placed in a homogeneous static magnetic field generated by permanent magnets. Two distinct alignment geometries were carried out: longitudinal alignment (MNPs with the magnetization aligned parallel to the tube axis, $M_L$) and transversal alignment (MNPs with the magnetization aligned perpendicular to the tube axis, $M_T$ ).

In both configurations, the samples were gelled *in situ* by temperature reduction to immobilize the aligned MNPs. All samples were prepared 24 hours before the US generation/detection experiments and stored sealed at 10°C to maintain stability.

Figure 1.a shows a schematic view of the relative orientation of the $M_L$ and $M_T$ aligned samples with respect to the direction of the applied EMF ($H_{EMF}$) supplied by the radiofrequency coil. Fig 1.b and 1.c respectively show images obtained by an optical microscope in which we can observe the high degree of alignment of the MNPs in the samples.

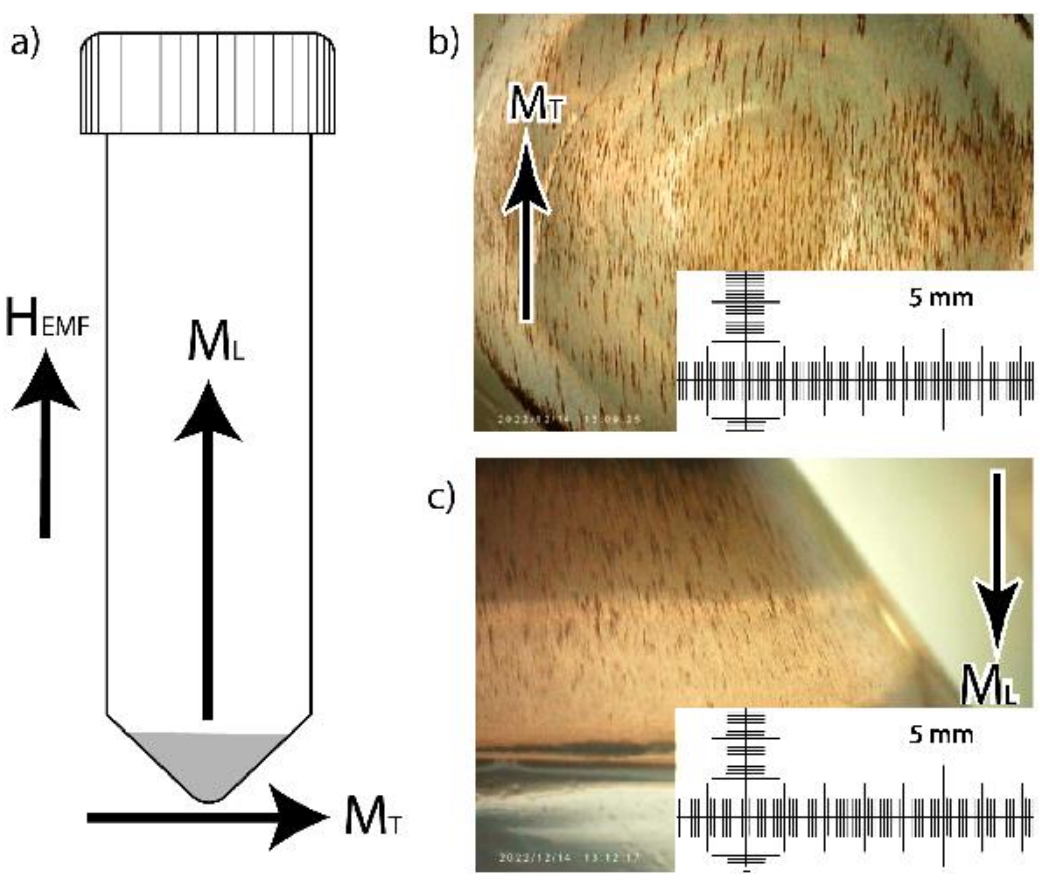

*Figure 1: a) Scheme of the 50 mL falcon loaded with a 250 μL solution of MNPs showing the direction of the applied EMF ($H_{EMF}$, along the falcon axis) and the direction of the magnetization of the aligned MNPs during the experiment. Optical microscopy images of magnetically jellified samples: b) MNPs transversally aligned, $M_T$, (perpendicular to the falcon axis) and c) MNPs longitudinally aligned, $M_L$, (along the falcon axis).*

The experimental setup is shown in Fig. 2. It consists of two main components: the EMF applicator (RF coil) and the US receptor. An aluminum support portal serves as a holder for the sample, which is contained in a 50 mL Falcon tube filled with distilled water, with the MNP sample at the bottom, as shown in Fig. 2. The Falcon tube is positioned on top of the RF coil that generates the EMF, powered by a CEIA Power Cube® generator and cooled by a TAE Evo® chiller. Above the Falcon, another support holds an Olympus® VS303 US transducer, which is inserted into the Falcon and submerged in the water, for the detection of the generated US.

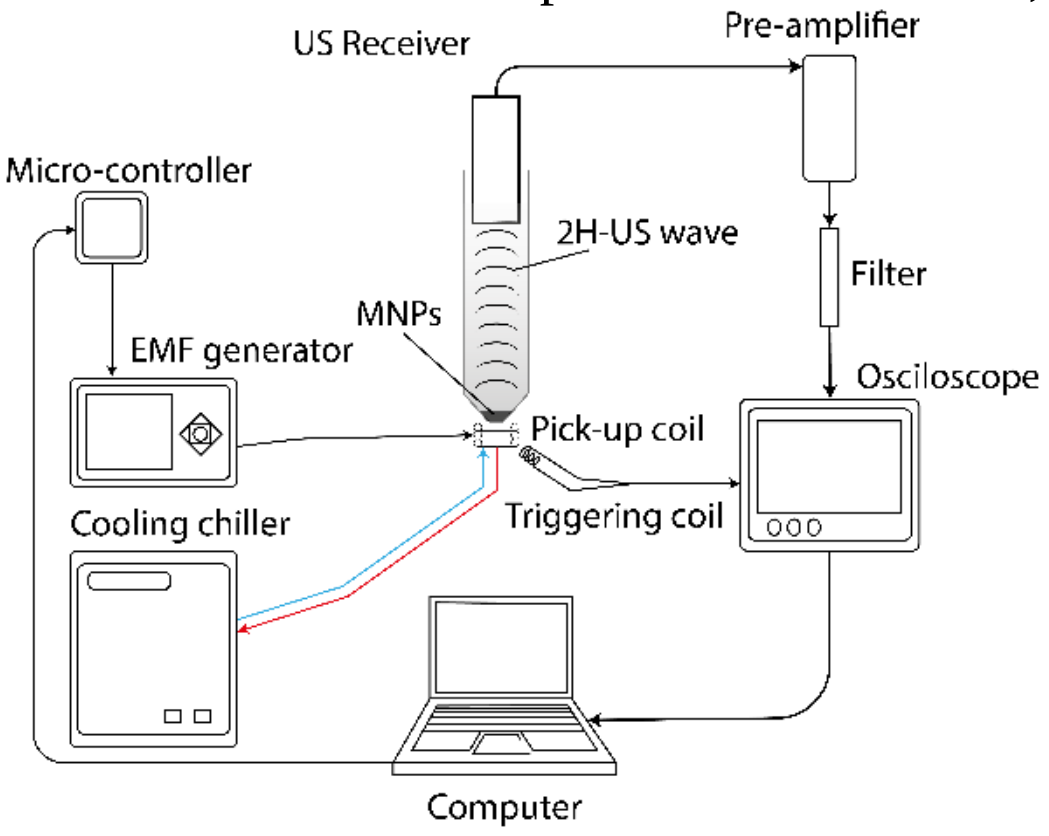

*Figure 2 Scheme of the experimental setup*

The detected acoustic signal is pre-amplified, filtered using an ONDA HFO-690 high pass filter (>2MHz), and analyzed using a Lecroy HDO4000A® oscilloscope. For the US reception the oscilloscope was synchronized with the signal induced in an auxiliary pick-up coil by the applied EMF. The applied EMF is controlled in duty cycles by a programmable M5Stack® microcontroller.

To ensure isothermal conditions during the experiments, we applied a duty cycle of 100 ms bursts of an EMF at 800kHz with an amplitude of 65mT. To avoid electromagnetic noise and monitor the temperature of the probe, we used an optic fiber sensor (H201 by Rugged Monitoring®). Experiments conducted on five samples demonstrated that temperature fluctuations remained below 1°C. We implemented a code in MATLAB (R2019b) software to analyze signals obtained from the oscilloscope.

In Fig. 3 we summarize the signal processing. Once the raw signals are obtained from the oscilloscope, they are processed in two ways to facilitate their further analysis. We applied the fast Fourier transform (FFT) to obtain the signal of the different harmonics, with a particular focus on the SH (Figure 3.b). The raw signal (Figure 3.a) is processed using a bandpass filter to eliminate frequencies far from the SH frequency. Arrival delay matches the acoustic path, confirming mechanical rather than electrical origin (Figure 3.c). This filtering method is helpful to reduce undesirable noise and enhance the quality of the signal. A Hilbert envelope and a smoothing filter are applied to the filtered signal to enhance the visualization of the distance at which the second harmonic US signal appears [13,14]. Finally, SH amplitude is statistically analyzed from measurements in ten samples as shown in the results of Figure 5.

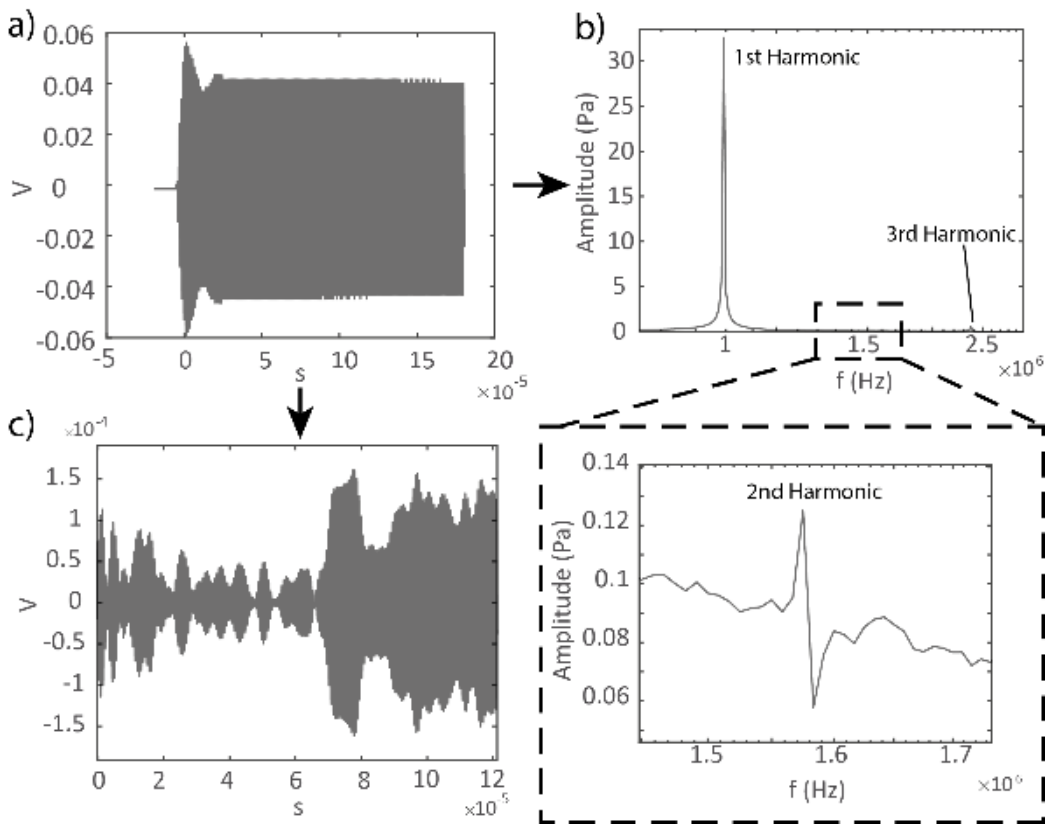

*Figure 3 US signal post processing. a) Raw signal b) FFT of the raw signal c) SH filtered signal.*

To confirm the magnetoacoustic nature of this signal, we performed the same experiment on a control hydrogel sample prepared with a non-magnetic alumina nanoparticle concentration similar to that of the MNP-hydrogel samples.

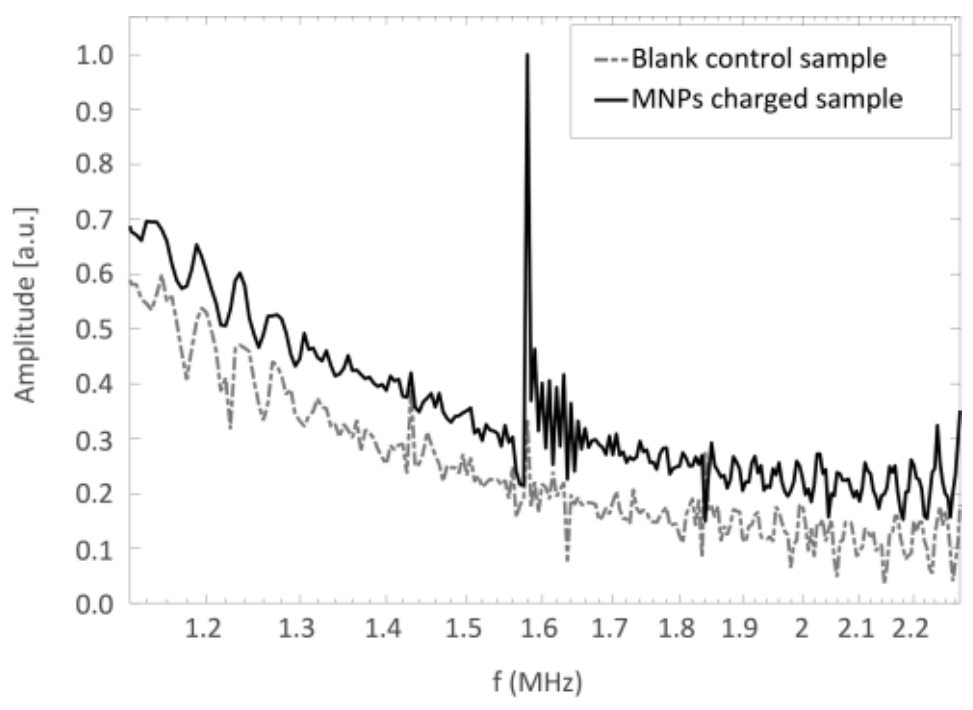

*Figure 4, comparison of the normalized signal obtained from hydrogel containing non-magnetic and magnetic nanoparticles.*

Figure 4 shows a comparison of the signal obtained from hydrogel containing non-magnetic aluminum oxide and MNPs. All SH amplitudes are referenced to the maximum detected signal in both samples. A clear peak is observed only in the FFT analysis of the signal obtained from the MNP samples, which corresponds to the 2H frequency. This clearly provides significant evidence of the magneto-acoustic interaction origin of this effect. Unlike previous optical or adiabatic demonstrations, this work presents, to our knowledge, the first observation of SH-US generation under strictly isothermal MHz RF excitation.

Based on this effect, we can use the SH signal to determine the allocation of MNPs by measuring the distance between the sample and the US transducer. We observed that the filtered signal at the second harmonic is time-shifted depending on this distance. These findings confirm the potential of MNPs as contrast agents in US imaging and suggest promising avenues for future research in this field that we will discuss further.

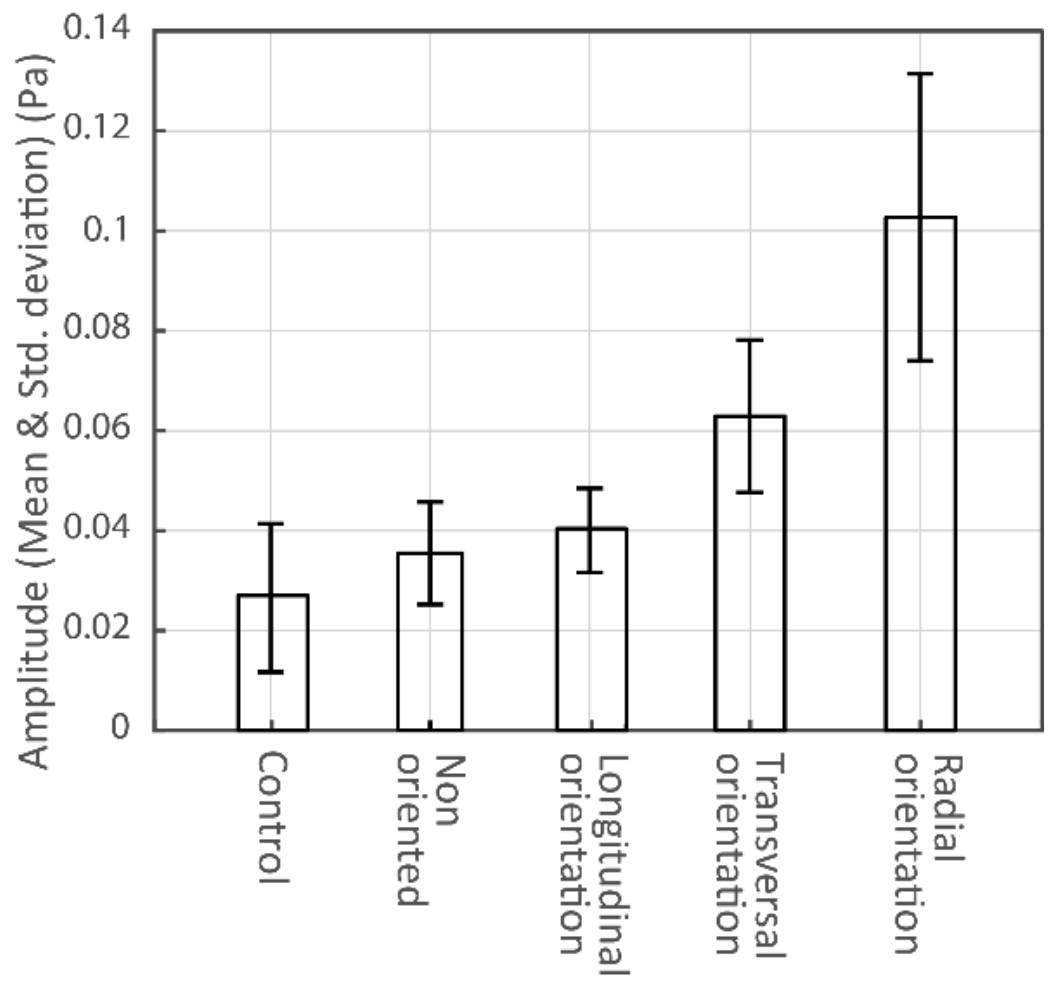

*Figure 5 Intensity of the SH emission for no MNP as control and MNP with and without orientation*

Different sets of hydrogel samples were measured: without MNPs, with non-aligned MNP magnetization and with MNP magnetization $M_L$, oriented along the direction of the applied EMF and with MNP magnetization oriented in the perpendicular direction, $M_T$. In this last case two situations were considered: in-plane magnetization parallel alignment and radial alignment. The results are shown in Figure 5. We clearly observed an enhancement of the SH signal of magnetically aligned samples with respect to the randomly magnetically oriented samples, as already was predicted by Carry et al. [7] but never observed. Moreover, our experiment showed that if the magnetization alignment of the MNP is perpendicular to the direction of the applied EMF ($M_T$ samples) the magnetoacoustic interaction is significantly enhanced. In particular the radial orientation in the $M_T$ samples favors the SH-US emission.

These results provide experimental verification of nonlinear magneto-mechanical coupling predicted by Carrey et al. at MHz frequencies under isothermal conditions, opening routes toward, but not limited to, biomedical applications. The detection of the SH-US is particularly relevant as is inherently free from electromagnetic interferences originated from the fundamental EMF excitation (first harmonic). We have demonstrated that the effect is strictly related with the presence of magnetoacoustic interaction. Our previous results demonstrated the cell death in MNP-loaded cells without temperature rise [5], here we observed that controlling the applied EMF by short pulses of radiofrequency fields we were able to irradiate MNPs keeping constant the temperature and detect the generation of US. The intensity is very low but could be enough to trigger biochemical mechanism at intracellular level to induce cell death. [15]

This study reveals, for the first time, that the magnetic alignment of nanoparticles significantly enhances the detection efficiency of the second harmonic (SH) signal, thereby offering a pathway for optimizing signal acquisition protocols. A further relevant outcome is the demonstrated potential for detecting MNPs within complex biological systems, such as cells and tissues. To address this, we developed a PZT-based scanner. This system provides distinct improvements over prior systems based on laser interferometry. The characteristics of this technique suggest improved feasibility for future clinical translation

## Author declaration section

The authors have no conflicts to disclose.

## Acknowledgements

This research was funded by the Ministry of Science and Innovation, Spain grant numbers PID2019-106947RA-C22 + PID2020 + DPI 2017 + P18 + EQ + IE. Project Gobierno de Aragón MAGNA E28-23R

M.R. Ibarra and G. Rus acknowledge to the University of Granada the support under the Program Visiting Scholar.

## Data availability statement

The data that support the findings of this study are available from the corresponding author upon reasonable request.

## References.